\documentstyle[epsf,11pt]{article}
\begin{document}
%
\title    {Masses of the physical mesons\break
             from an effective QCD--Hamiltonian}
\author{Hans-Christian Pauli and J\"org Merkel \\ 
        Max-Planck-Institut f\"ur Kernphysik \\
        D-69029 Heidelberg \\ }
\date{22 August1996}
\maketitle

\begin{abstract}
The front form Hamiltonian for quantum chromodynamics, 
reduced to an effective Hamiltonian acting only in the 
$q\bar q$ space, is solved approximately. After coordinate
transformation to usual  momentum space and Fourier
transformation to configuration space a second order
differential equation is derived. This retarded  Schr\"odinger
equation is solved by variational methods and semi-analytical 
expressions for the masses of
all 30 pseudoscalar and vector mesons are derived.
In view of the direct relation to quantum chromdynamics 
without free parameter, the agreement with experiment 
is remarkable, but the approximation scheme 
is not adequate for the mesons with one up or down quark.
The crucial point is the use of a running  coupling constant 
$\alpha_s(Q^2)$, in a manner similar but 
not equal to the one of Richardson in the equal usual-time 
quantization. Its value is fixed at the 
Z mass and the 5 flavor quark masses are determined by 
a fit to the vector meson quarkonia.
\end{abstract}
\vfill 
\noindent \\
Preprint MPIH-V34-1996 \\
\\
Submitted to Physical Review D (1996) \\
PACS-Index: 12.40.Yx, 11.10.Ef, 11.15.Tk \\
\\
Revision of MPIH-V45-1995  (2 December 1995) \\
Internet: pauli@zooey.mpi-hd.mpg.de \\

\section{Introduction and motivation}

One of the most outstanding tasks in strong interaction 
physics is to calculate the spectrum and the wavefunctions 
of physical hadrons from quantum chromodynamics (QCD).
Discretized light-cone quantization (DLCQ) \cite{PaB85} 
has precisely this goal. Its three major aspects are: 
(1) a rejuvenation of the Hamiltonian approach, 
(2) a denumerable Hilbert space of plane waves, and
(3) Dirac's front form of Hamiltonian Dynamics.
In the {\em front form} \cite{Dir49}, or in 
{\em light-cone quantization} \cite{Wei66},
one quantizes at equal `light-cone time' $x^+=t+z$, 
as opposed to the conventional {\em instant form} 
where one quantizes at equal usual time $t$.
As reviewed in \cite{BrP91}, the front form has unique 
features, among them: The vacuum is simple, 
or at least simplier than in the instant form, and 
the relativistic wavefunctions transform trivially under 
certain boosts \cite{Dir49,BrP91}.
Both are in stark contrast to the conventional instant form.
Over the years, the light-cone approach \cite{Gla95} 
has made much progress. Calculations \cite{BrP91} 
agree with other methods particularly lattice gauge theory.
Zero modes of the fields  can be important carriers 
of quantum structures \cite{HKS92}, particularly  of 
those of the vacuum \cite{DKB94,PKP95}. 
Dimensionally reduced models \cite{DKB94,PKP95} 
provide much insight into the structure of possible 
solutions to QCD. But chiral aspects are not yet 
understood, and non-perturbative renormalization 
remains a challenge \cite{PHW90,WWH94} 
as for any Hamiltonian approach. 

But despite the many successes of light-cone Hamiltonian
methods one misses the contact to phenomenology beyond
the perturbative regime. We believe that  more QCD-inspired 
approaches are needed, work like for example 
\cite{BSW87,mit89} or \cite{BrS94,Sch94}, 
where the formalism is related to the experiment. 
The present work is of this type. 

Right from the outset when applying DLCQ to gauge theory 
in 3+1 dimensions \cite{tbp91,KPW92,sol93} 
it was clear that one should need an effective Hamiltonian. 
In \cite{KPW92} an integral equation in the light-cone 
momenta was solved  numerically, which was derived  by 
procedures similar to those  of Tamm 
and Dancoff \cite{Tam45,Dan50}, and a non-integrable
singularity was removed by an {\it ad hoc} assumption.
But recently \cite{Pau93} the method of effective interactions
was generalized to avoid the usual truncation in the particle
number \cite{Pau96}.  As it turns out, one can assemble all
many-body aspects into a vertex function which bears great 
similarity with the running coupling constant. 

One wonders: How can such a simple structure
account for the {\em spectra
and wavefunctions of all scalar and vector mesons}?
Is this not too much of a claim? On the other hand
the effective Hamiltonian has been derived \cite{Pau96} 
from the QCD-Lagrangian without condition on the 
coupling constant or on the mass of the constituents.
One way of checking this is to compare to experiment, 
and this shall be done in this work very roughly and
preliminarily. Lacking the running  coupling constant 
going with the theory \cite{Pau96}, one can replace 
it by one of its current phenomenological versions
\cite{Ric79,Cor82,GDH93}.
The present work applies the one
of  Richardson \cite{Ric79}. It interpolates smoothly 
between  asymptotic  freedom \cite{Pol73,GrW73}
and infrared slavery. After that, one has no freedom 
in the theory and no  adjustable parameters. 
Since the quark masses cannot be determined from 
independent measurements, they must be determined
self-consistently from a fit to some of the meson masses. 
This in itself is not trivial, except when having analytical 
expressions.

In particular, a coordinate transformation from front 
to instant-form coordinates is performed in Section~3.
Apart from a more transparent interpretation, this way of
writing down the integral equation has certain advantages
in performing the calculations. No assumptions will be made
in this section: All manipulations are straightforward and 
fully equivalent to the front-form formulation.
In Section~4, the bound-state equation is approximated 
semi-relativistically which allows for Fourier transforming 
the momentum-space integral equation into a
configuration-space Schr\"odinger type equation. The so
obtained Hamiltonian is reduced in Section~5 to a minimal
number of terms  (Coulomb plus linear potential plus one 
spin-dependent term distinguishing between singlet and 
triplet) and diagonalized approximately by a variational 
method.

The masses of all pseudoscalar and vector mesons in 
Section~6 are thus semi-analytic and approximate 
solutions to a second order differential equation 
in configuration space. In comparison with the empirical 
masses \cite {PDG94}, they are not much worse than those 
of potential models \cite{QuR79,GoI85,lsg91,mns93,cad94}, 
or predictions from heavy quark symmetry \cite{Neu93}, 
or even predictions based on lattice gauge calculations 
\cite{Mac93,BCS94,Wei94}. In view of the direct link to 
QCD \cite{Pau96} and the simplicity  this
should be regarded as considerable progress in the front-form
approach.

But there is a potential danger in such an endeavor. 
The present work is motivated by the question whether 
the simple structures to be displayed can describe 
experiments {\em at all}. Obviously they can, but it should 
be emphasized that numerically accurate solutions need 
another effort. This is currently being attempted \cite{TrP96} 
and has a different objective than to develop models
designed to reproduce the data. 
\section{The effective Hamiltonian for QCD}\label{sec:2}
\begin{figure} [t]
\begin{minipage}[t]{80mm} \makebox[0mm]{}
\epsfxsize=80mm\epsfbox{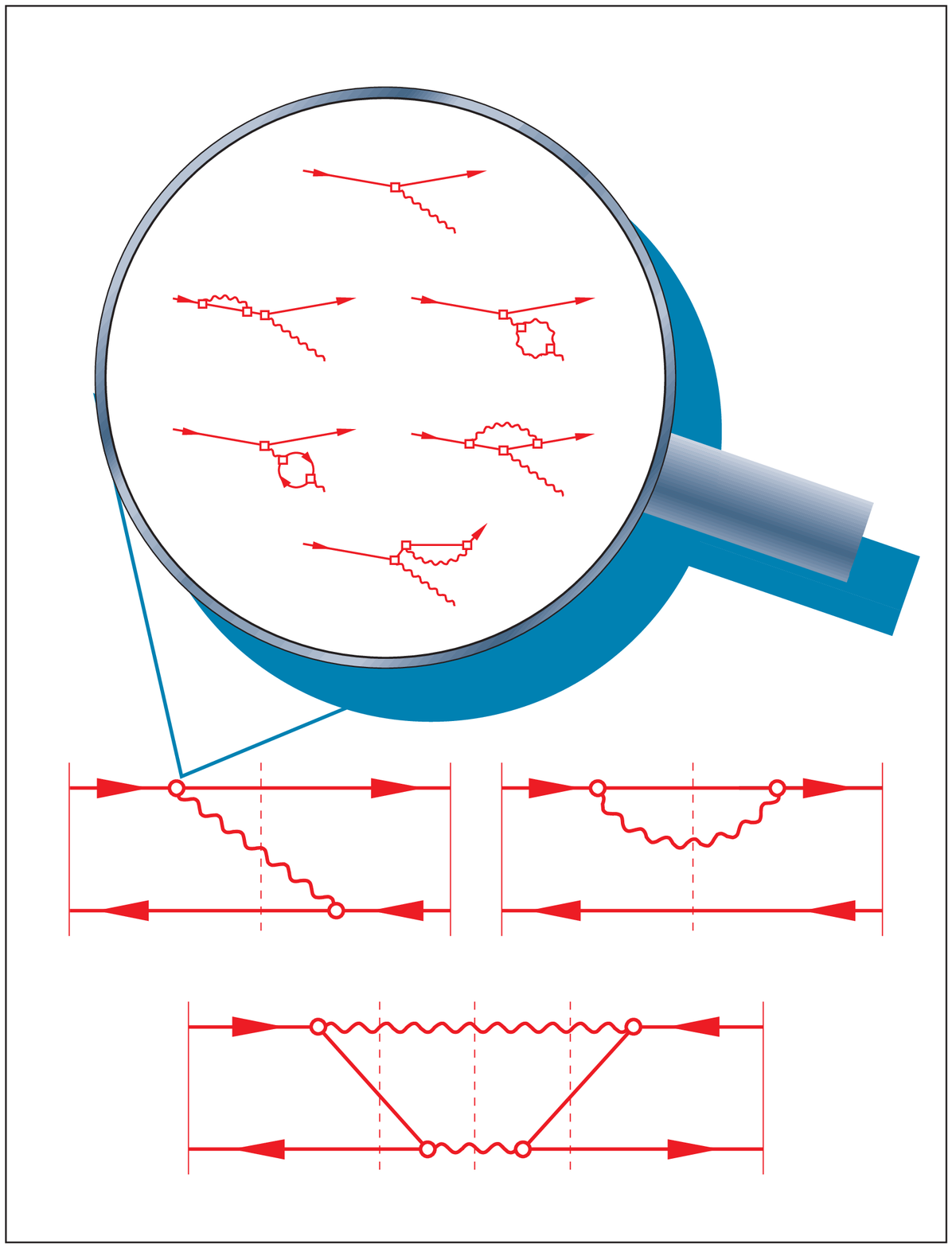}
\caption{\label{fig:1}\sl
The effective interaction in the $q\bar q$ sector.
By exchanging `effective gluons', 
a single quark state with four-momentum $k_1$ and 
spin projection $\lambda_1$ is scattered into the quark 
state ($k_1^\prime,\lambda_1^\prime$).
Correspondingly, the antiquark is scattered from
($k_2,\lambda_2$) to ($k_2^\prime,\lambda_2^\prime$).
} \vfill \end{minipage}
\hfill
\begin{minipage}[t]{80mm} \makebox[0mm]{}
\epsfxsize=80mm\epsfbox{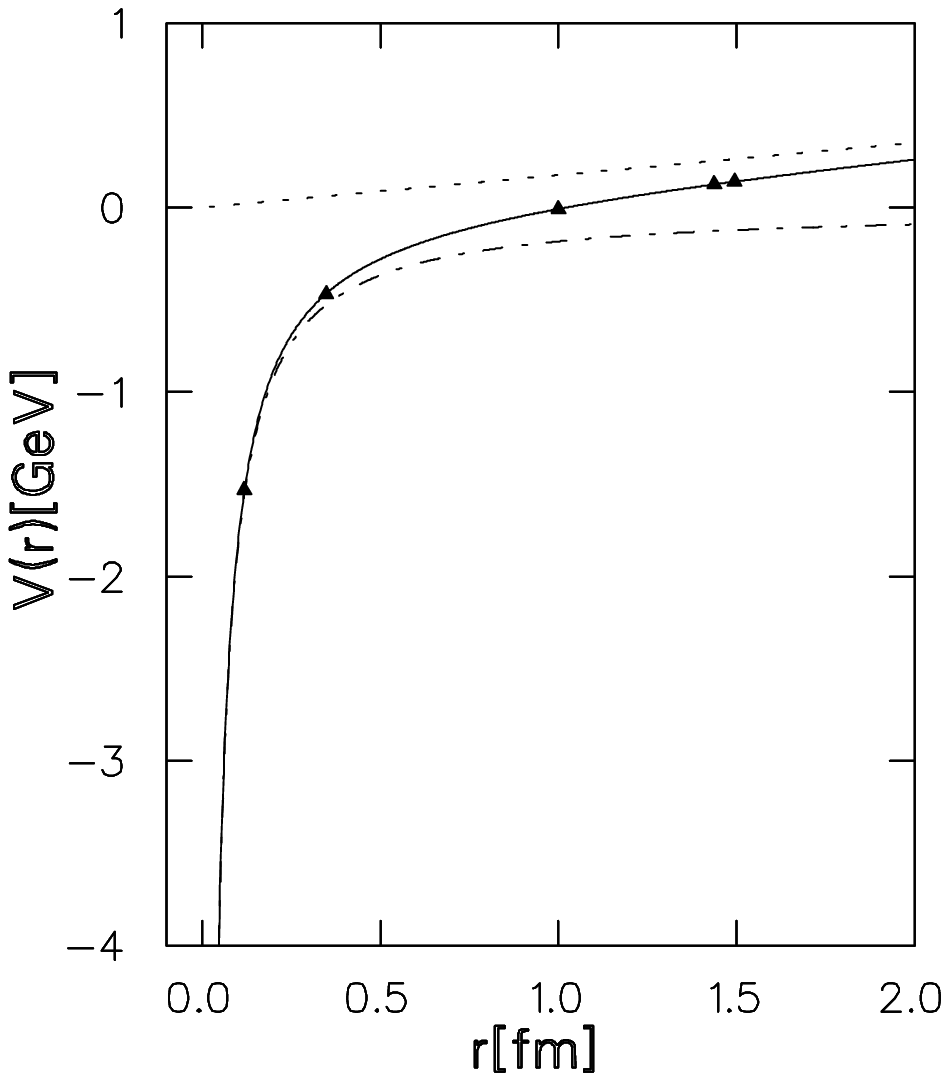}
\caption {\label{fig:2}\sl
The quark-antiquark potential $V(r)$ as given
in Eq.(\protect\ref{eq:11}) is plotted versus the
relative distance $r$. 
The Coulomb 
and the confining potential  are indicated.
Some of the heavier meson masses are
inserted to provide a scale.  
} \end{minipage}
\end{figure}

In Discretized Light-Cone Quantization (DLCQ) one 
seeks to solve the eigenvalue problem 
\begin{equation}
      H _{\rm LC} |\Psi_b\rangle = M_b ^2 |\Psi_b\rangle
\label {eq:1} \end{equation}
for a field theory. The `light-cone Hamiltonian' 
$ H_{\rm LC} \equiv  P^{\mu}P_{\mu} $ \cite{BrP91} is the 
Lorentz invariant contraction of the energy-momentum 
four-vector  $P^\mu$ and has the dimension  
$<\!\!mass\!\!>^2$.  The eigenvalues $M _b^2$ are 
interpreted  as the square of the invariant mass  of state $b$. 
Working in momentum representation, the three space-like 
components of $P^\mu$ are diagonal operators, with
eigenvalues $ P ^+ = \sum _j k ^+ _j $ and 
$ \vec P _{\!\bot} = \sum _j \vec k _{\!\bot _j }  $. 
The sum runs over all particles in a Fock state.
Each particle has a four-momentum denoted by 
$ k ^\mu _j = (k ^+ _j , \vec k _{\!\bot _j }, k ^- _j) $ 
and sits on its mass shell $ ( k ^\mu k _\mu) _j = m _j ^2$.
The time-like component, the Hamiltonian proper $ P ^-$, 
is a complicated and off-diagonal operator acting
in Fock space. Its matrix elements are tabulated in
\cite{BrP91}.  Based on the boost-properties of light-cone 
operators \cite{BrP91} one can transform to a frame where 
$ \vec P _{\!\bot} = 0$,  thus $ P ^\mu P_\mu = P ^+ P ^- $.
Since $ P ^+$ is diagonal, the diagonalization of 
$P ^-$ and of $ H _{\rm LC} $ amounts to the same.
The Hilbert space for diagonalizing $P ^-$ is spanned by 
all Fock states which have given eigenvalues of 
$P ^+ $  and $ \vec P _{\!\bot} = (0, 0) $ and can be 
arranged into sectors according to the particle number 
like $q\bar q $, $q\bar q \,g$, or $q\bar q \, q\bar q $. 
For any fixed value of the harmonic resolution 
$K=2LP ^+/\pi$,
the  Hamiltonian matrix in Eq.(\ref{eq:1}) is finite and in 
principle could be diagonalized numerically \cite{PaB85}. 
Details can be found in the literature \cite{BrP91,Pau93,Pau96}. 

DLCQ is quite useful to cleanly phrase the problem, but to
do calculations particularly in 3+1 dimensions one has to
develop effective Hamiltonians. Fock space truncation in
conjunction with perturbation theory in the manner of 
Tamm and Dancoff \cite {Tam45,Dan50} is unsatisfactory, 
because one has to resort to {\it ad hoc} prescriptions to
make things work \cite{KPW92}. These drawbacks can 
be avoided by the method of iterated resolvents 
\cite{Pau93,Pau96}. It turns out possible to convert the 
many-body matrix Eq.(\ref{eq:1})  into a well defined
two-body equation with an effective interaction  
acting only in the $q\bar q$-space, {\it i.e.}  
$H_{\rm eff}\vert\psi_b\rangle=M_b^2\vert\psi_b\rangle$.
In the {\it continuum limit} one has to solve the  
integral equation 
\begin{equation} 
        M_b^2 
        \langle x,\vec k_{\!\perp}; \lambda_1,
        \lambda_2 \vert \psi _b\rangle 
        = 
        \sum _{ \lambda_1^\prime,\lambda_2^\prime }
        \! \int\!\! dx^\prime d^2 \vec k_{\!\perp}^\prime
        \,\langle x,\vec k_{\!\perp};\lambda_1,\lambda_2
        \vert H_{\rm eff}\vert x^\prime,\vec k_{\!\perp}^\prime; 
        \lambda_1^\prime,\lambda_2^\prime \rangle \,
        \langle x^\prime,\vec k_{\!\perp}^\prime; 
        \lambda_1^\prime,\lambda_2^\prime 
        \vert \psi _b\rangle 
.      \label {eq:2} \end{equation}
The bras and kets refer to $q\bar q$ Fock states which
can be made invariant under $SU(N)$, 
\begin{equation}
        \vert  x,\vec k_{\!\perp}; \lambda_1,\lambda_2  
        \rangle  ={1\over \sqrt{n_c}} \sum _{c=1} ^{n_c}
       b^\dagger_c (k_1,\lambda_1) 
       d^\dagger_c  (k_2,\lambda_2) \vert 0 \rangle 
.      \end{equation}
Goal of the calculation are  the momentum-space
wavefunctions  $\langle x,\vec k_{\!\perp};  
\lambda_1,\lambda_2    \vert \psi _b\rangle $.
They are the probability amplitudes for finding the quark 
with helicity projection $\lambda_1$, longitudinal momentum 
fraction $x\equiv k_1^+/P^+$ and transversal momentum 
$ \vec k _{\!\bot}$, and correspondingly the antiquark with 
$\lambda_2$, $1-x$ and $-\vec k_{\!\bot}$.  
The effective interaction as diagrammatically displayed 
in Figure~\ref{fig:1} is a sum of three terms: 
The first two diagrams are kind of a one-gluon exchange
and  describe the flavor conserving part of the effective 
interaction, while the last graph due to the two-gluon 
annihilation can change  the flavor. 
In the present work we deal only with the first of them.
The kernel of the integral equation (\ref{eq:2}) has  a
diagonal `kinetic' and an off-diagonal `interaction' energy, 
{\it i.e.} 
\begin {eqnarray} 
        M_b^2 \,
        \langle x,\vec k_{\!\perp}; \lambda_1,\lambda_2  
        \vert \psi _b\rangle 
        = \left[ {m_1 + \vec k_{\!\perp}^2 \over x } 
        + {m_2 + \vec k_{\!\perp}^2 \over 1-x } \right]
        \langle x,\vec k_{\!\perp}; \lambda_1,\lambda_2
        \vert \psi _b\rangle 
\nonumber\\
        - {1\over 4\pi^2}
        \sum _{ \lambda_1^\prime,\lambda_2^\prime }
        \! \int\!\! dx^\prime d^2 \vec k_{\!\perp}^\prime
        \,\Theta(x^\prime,\vec k_{\!\perp}^\prime)
        \,{\beta(Q) \over Q  ^2} 
        \,{\langle\lambda_1,\lambda_2
        \vert S(Q) \vert 
        \lambda_1^\prime,\lambda_2^\prime \rangle 
        \over \sqrt{x(1 - x) x^\prime (1- x ^\prime)} } \,
        \langle x^\prime,\vec k_{\!\perp}^\prime; 
        \lambda_1^\prime,\lambda_2^\prime 
        \vert \psi _b\rangle 
.      \label {eq:4} \end {eqnarray}
The most important factors are the four-momentum transfer 
\begin {equation}
        Q ^2 = - (k_1 - k _1^\prime)^2  = - (k_2 - k _2^\prime)^2
       \label {eq:5} \end {equation}  
and the vertex function  $r(Q,\Lambda)$ which 
likes to combine with the coupling constant $g$ to become
\begin{equation}
        \beta(Q) = {n_c^2-1\over 2n_c}
        \  {g^2\over 4\pi \hbar c}  \  r^2(Q,\Lambda)
        \equiv {4\over3} \alpha_s(Q)
,       \end{equation} 
the like-to-be `running coupling constant'.
For QED, this factor reduces to the fine structure constant
$\beta = \alpha\sim 1/137$.  
The spinor factor $S(Q)$ represents the 
familiar current-current coupling 
\begin{eqnarray} 
        \langle \lambda_1,\lambda_2\vert S(Q) \vert 
        \lambda_1^\prime,\lambda_2^\prime \rangle 
        = 
        \left[ \overline u (k_1,\lambda_1) 
        \,\gamma^\mu\,
        u(k_1^\prime,\lambda_1^\prime)\right] 
        \left[ \overline u (k_2,\lambda_2) 
        \,\gamma_\mu\,
        u(k_2^\prime,\lambda_2^\prime)\right] 
.       \label{eq:7}\end{eqnarray} 
The cut-off function 
$\Theta(x^\prime,\vec k_{\!\perp}^\prime)$ restricts 
integration in line with Lepage-Brodsky 
regularization \cite{BrP91}
\begin{equation}
        \Theta(x,\vec k_{\!\perp}):\qquad\qquad
        {m_1^2 +\vec k _{\!\bot}^{\,2}\over x } +
        {m_2^2 +\vec k _{\!\bot}^{\,2}\over 1-x }
        \leq (m_1+m_2)^2 + \Lambda ^2 
.      \label {eq:8} \end{equation}
The mass scale $\Lambda$ can be chosen freely.

Despite having been derived in the light-cone gauge 
$A^+ = 0$,  the effective interaction is {\em manifestly 
gauge invariant}, depending only on the quark currents. 
The instantaneous interaction has cancelled exactly 
against other gauge-variant terms, see
\cite{tbp91,KPW92,Pau96}.  
Since one works in the front form, it is also 
{\em frame and boost invariant}.  
Explicit calculations for QED \cite{KPW92,TrP96} 
are numerically very stable, and reproduce quantitatively 
the Bohr spectra and the fine and hyperfine structure.  

The vertex function hidden in the like-to-be running coupling 
constant of Eq.(\ref{eq:5}) has the same perturbative series 
expansion as the running coupling constant \cite{Pau96}  
which is indicated in an artists way in Figure~\ref{fig:1}.  
What is missing, thus far, is a renormalization group analysis 
of the formal expressions. 
In the absence of that, we are interested in consequences
of Eq.(\ref{eq:5}). How can it be that such a simple
expression accounts for hadronic phenomena?
What are the invariant masses of the pseudoscalar and vector 
mesons, using such an interaction? How far does one get 
with analytical procedures, and in particular where does the
approach go wrong? 

\begin{figure} [t]
\begin{minipage}[t]{80mm} \makebox[0mm]{}
\epsfxsize=80mm\epsfbox{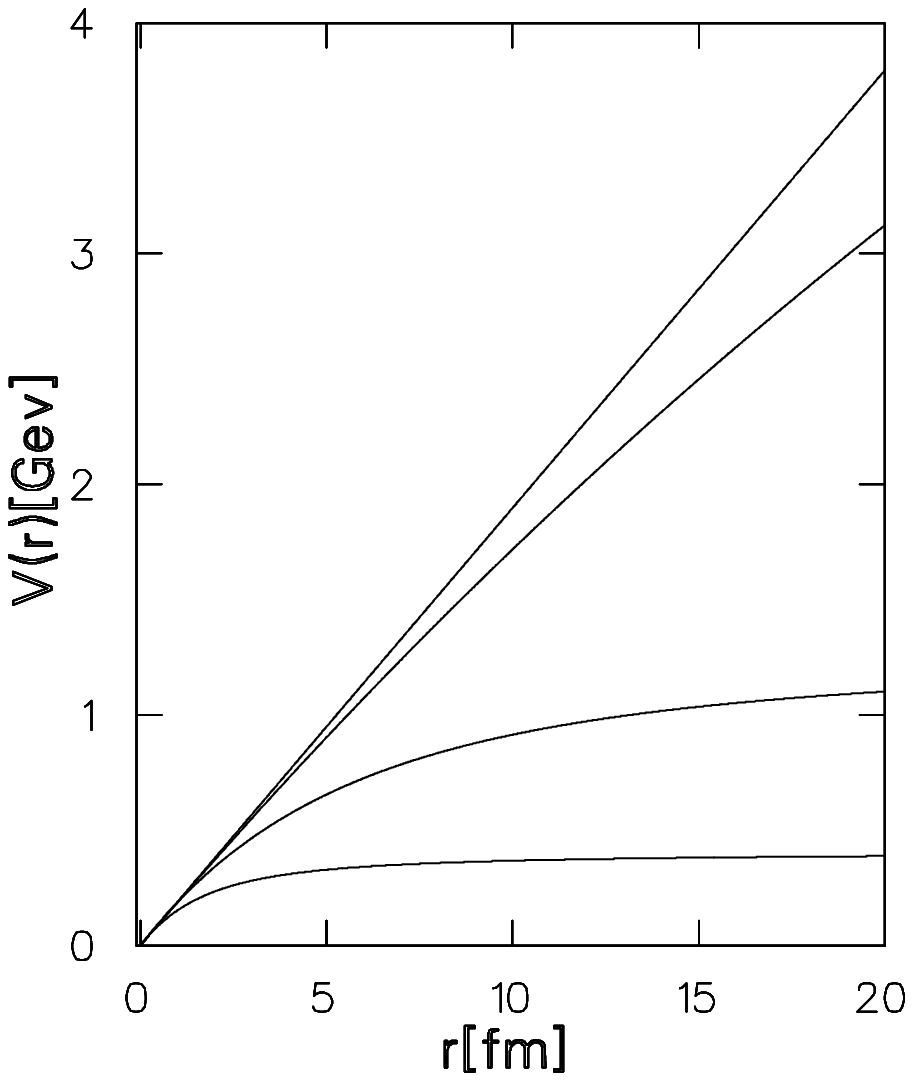}
\caption {\label{fig:4}\sl
The confining potential as function of the 
parameter $a$. Values are from top to bottom:
$a=1$, $a=1.0005$, $a=1.01$, and 
$a=\protect\sqrt{e}\simeq 1.65$. 
} \vfill \end{minipage}
\hfill
\begin{minipage}[t]{80mm} \makebox[0mm]{}
\epsfxsize=80mm\epsfbox{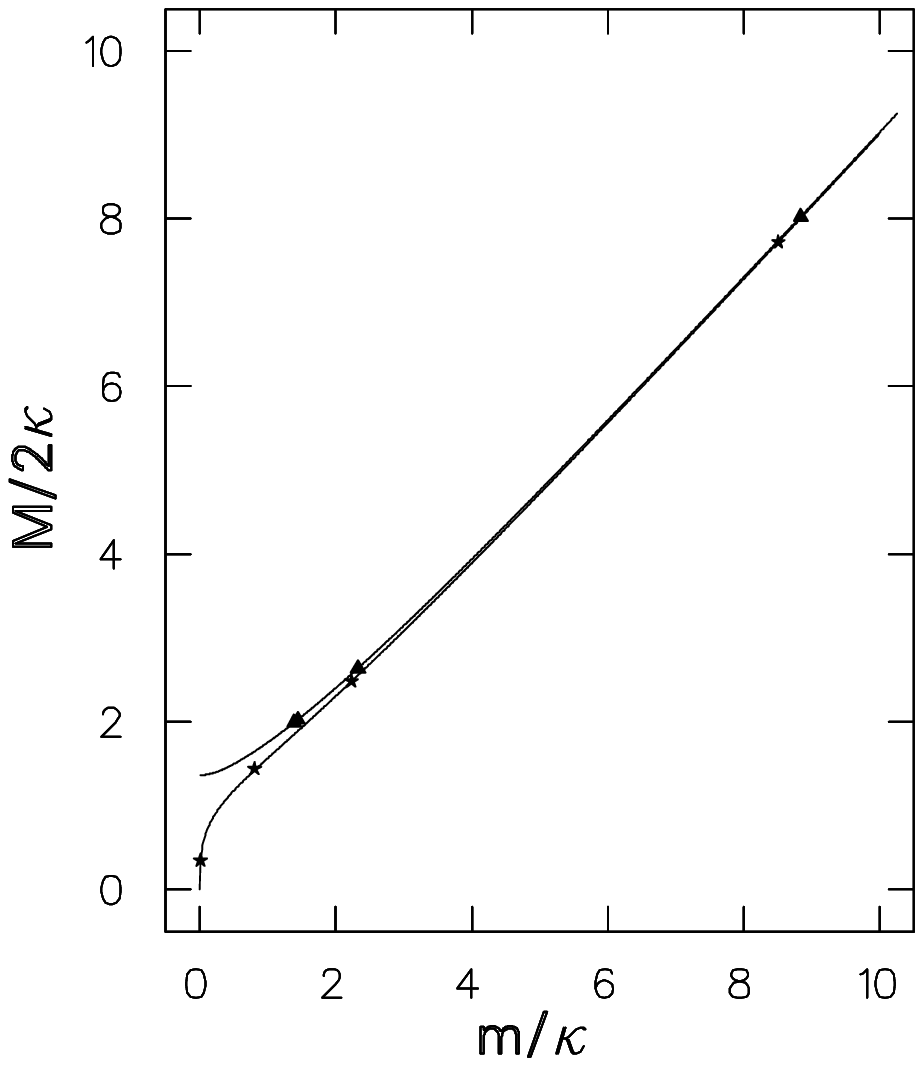}
\caption{\label{fig:3}\sl
Bound states of $q\bar q$-pairs {\it versus} quark masses. 
All masses are given in units of the QCD-scale $\kappa$. 
The upper curve refers to the triplet ($S_e=2$), the lower 
to the singlet ($S_e=0$). The masses of some vector 
mesons ($\rho^0,\omega,\phi,J/\psi$) are marked by a
($\triangle$), those of the some pseudoscalars 
($\pi^0,\eta, \eta',\eta _c$) by a ($\star$). 
} \vfill \end{minipage}
\end{figure}

Lacking an exact expression for $\alpha_s( Q ^2)$, 
one can resort to reasonable parametrizations 
\cite{Ric79,Cor82,GDH93}. In the sequel, we shall content 
ourselves with the form of Richardson \cite{Ric79}, 
\begin{equation}
     \alpha _s ( Q ^2) = \frac{12\pi}{27}\, 
     \frac{1}{\ln(a ^2+ Q ^2/\kappa ^2 )}
\ .  \label{eq:9}\end{equation}
At least, this form interpolates smoothly between asymptotic
freedom  \cite {Pol73,GrW73} and infrared slavery.
In the original work, the the parameter $a$ was set to have
the value $a=1$ and $\kappa$ was kept as free parameter 
to be determined by the spectra. Here, we take
the value of $\alpha _s(M_Z)=0.1134 \pm 0.0035$ 
as measured at the Z-mass \cite{PDG94} to fix 
\begin {equation}
     \kappa = 193\, {\rm MeV}
\ . \end {equation}  
Its Fourier transform \cite{Ric79} generates two terms, 
see also below, 
\begin {equation}
        V(\vec x\,) = {-1\over 2 \pi ^2} 
        \int\!\! d ^3 \vec q
        \ {\rm e} ^{ i\vec q \cdot \vec x }
        \ {\alpha_s (\vec q ^{\, 2}) \over \vec q ^{\, 2} }
  =   {8\pi\over27} \Big( - \frac{1}{r} + \kappa^2\, r \Big)
\ , \label {eq:11} \end {equation}
a (strong) Coulomb and a linearly rising potential, 
as plotted in Figure~\ref{fig:2} versus $r=\vert\vec x\vert$. 
The linearity of the confining potential is a consequence
of $a=1$, as used in \cite{Ric79} and throughout the 
present work. If one varies $a$ one gets  the curves 
displayed in Figure~\ref{fig:4}. It is taken from
\cite{Mer94}.  Here, we do not want to  keep $a$
as a free parameter. For one reason, we refuse to speculate
at this point whether the potential is strictly confining or not. 
For the other reason, the results to be displayed below are not
very sensitive to  large  distances since the wavefunctions
decay rapidly. Last but not least, one has to await the
renomalization group analysis of the $\alpha_s(Q)$ 
which truly comes with the theory \cite{Pau96}.

The flavor quark masses are then the only free parameters
of the approach. Of course, they are subject to be 
determined  consistently by experiment. 
Natural candidates are the masses of the pseudoscalar 
($0^-$) and vector  mesons ($ 1 ^- $). 
Since the flavor quark masses potentially range from 
a few MeV up to some 100 GeV, see for instance 
Figure~\ref{fig:3}, one faces two problems: 
(1) By good reasons, the numerical solutions of the
integral equation have been restricted thus far to
systems with equal masses of the constituents like 
positronium \cite{KPW92,TrP96}. The wave function is
then peaked at $x=m_1/(m_1+m_2)=1/2$. For very
asymmetric systems this will be a problem. To avoid that, 
we shall identically rewrite the integral equation 
in the next section in terms of instant-form variables,
which are somewhat easier to deal with.
(2) Shall one really perform a calculation of  similar 
complexity as in preceeding work \cite{KPW92,TrP96}
for any given set  of  $m _1$ and $m _2$ when intending  
to fit them to the meson masses, or shall one 
aim for a quasi-analytic but approximate solution? 
In view of the preliminary character of the present study, 
we have opted for the second. At least, this
will pave the way  for a future and improved solution.

In the sequel, we shall replace $H_{\rm eff}$
by the operator
\begin {equation}
       H =  {1\over 2(m_1+m_2)}
       \left(H_{\rm eff} - (m_1+m_2)^2\right)
\ . \label {eq:12} \end {equation}
It differs from $H_{\rm eff}$ by an additive constant
and an overall  scale, which both are Lorentz scalars.
Both $H$ and its eigenvalue $E$ have the dimension of a
$ \langle mass \rangle$ and have much in common with 
the non-relativistic Hamiltonian and the binding energy 
$E$, as we shall see.
As compared to an instant-form Hamiltonian, however, 
the main advantages of the front-form $H_{\rm eff}$ as 
given in Eq.(\ref{eq:5})  prevail, namely the additivity of 
the interaction  and the Lorentz invariance of the
eigenvalues.~--- The  rest of this work is a simple and 
straightforward evaluation.

\section{Transforming variables from the front to the 
instant form} \label{sec:3}

The single-particle four-momenta  can be parametrized 
either in the front form, 
$ k ^\mu _1 = (k^+_1,  \vec k _{\!\bot},k^-_1)$ and
$ k ^\mu _2 = (k^+_2, -\vec k _{\!\bot},k^-_2)$, 
or in the instant form, 
\begin {equation} 
     k ^\mu _1 
   = ( k ^0 _1 ,  \vec k _{\!\bot} , k _z)
   = ( E _1 , \vec k )
, \quad{\rm and} \quad
     k ^\mu _2 
   = ( k ^0 _2 , -\vec k _{\!\bot} , -k _z)
   = ( E _2 , - \vec k )
\ . \label {eq:13} \end {equation} 
Since $k_i^\mu k _{i,\mu} = m_i^2$, the time-like
components are functions  of the space-like components 
\begin {equation} 
     k ^-_i = 
    {m _i^2 + \vec k _{\!\bot}^{\;2} \over k_i^+} 
, \quad{\rm or} \quad
     E_i=\sqrt{m_i^2+\vec k_{\!\bot}^{\, 2}+\vec k_z^{\, 2}}
            =\sqrt{m_i^2+\vec k^{\, 2}}
\ . \end {equation} 
The transformation function between $x$ and $k _z$ 
is obtained straightforwardly from $P^+$, {\it i.e.}
\begin {equation} 
     x = x (k _z) = {k_z+E_1 \over E_1+E_2}
\ . \label {eq:15} \end {equation} 
The front form integral equation is boost and frame 
invariant, and therefore can be solved also in 
the center of mass frame,  
where the total momentum $\vec P$  vanishes.
Changing  integration variables, 
Eqs.(\ref{eq:4}) in conjunction with Eq.(\ref{eq:12}) 
can thus be rewritten identically as
\begin {equation} 
        E\,\langle\vec k\vert\psi\rangle 
        = T(\vec k \,)\,\langle\vec k\vert\psi\rangle 
        +  \int \! d^3 \vec k ^\prime
        \,\langle\vec k\vert   U
        \vert \vec k^{\,\prime} \rangle
        \,\langle\vec k^{\,\prime}\vert\psi\rangle 
.       \label{eq:16}\end {equation}
For simplicity the explicit summation over the 
helicities is suppressed.
Contrary to Eq.(\ref{eq:4}) all three integration variables
have now the same support. The kinetic energy 
\begin{eqnarray} 
     T (\vec k) \equiv
     {1\over 2(m_1 + m_2)} \left( 
     {m_1^2 + \vec k _{\!\bot}^{\;2} \over x} +
     {m_2^2 + \vec k _{\!\bot}^{\;2} \over 1-x} - 
     (m_1 + m_2)^2 \right) =
     {(E_1 + E_2)^2 - (m_1 + m_2 )^2 
     \over 2(m_1 + m_2)} 
\end{eqnarray}
becomes the familiar expression with the reduced mass 
$m_r$, for sufficiently small momenta, 
\begin{eqnarray} 
       T (\vec k) = {\vec k^{\,2} \over 2 m_r}
        \ ,\qquad{\rm with}\quad
       {1\over m_r} = {1\over m_1} + {1\over m_2}
\ .   \end{eqnarray}
Nevertheless, there are explicit residues from the front form.
The Fock state $ \vert x , \vec k _{\!\bot} \rangle $
has the same $ P ^+ $ as
$ \vert x' , \vec k _{\!\bot}' \rangle $.
Expressed in instant form variables $ P ^+ = P ^0 + P ^3$. 
Since $ P ^3 = k _{z,1} + k _{z,2} = 0 $  
one is left with $ P ^0 = P {^0} ' $, or explicitly
\begin {equation} 
       E _1 + E _2 = E _1 ' + E _2 ' 
       \ , \qquad {\rm or} \qquad 
       \vec k ^2 = \vec k ^{\prime\, 2} 
\label{eq:19} \end {equation} 
for every matrix element. 
Obviously, the interaction kernel in
Eq.(\ref{eq:16}) cannot change the size of $ \vec k $, 
it only changes its direction.  
This is a source of great simplification. 
For example, the four-momentum transfer is always 
identical with the three-momentum transfer  
\begin {equation} 
       Q^2 = - (k_1 - k'_1)^\mu (k_1 - k'_1)_\mu =
       (\vec k_1 - \vec k'_1) ^2 - ( E_1 - E'_1) ^2
        = \vec q ^{\;2} 
, \end {equation} 
and the three-momentum transfer and its mean, 
\begin {equation} 
      \vec q    = \vec k - \vec k '
      \qquad{\rm and} \quad
      \vec p = {1\over2} (\vec k + \vec k ' )
,  \end {equation} 
respectively, are always orthogonal: 
\begin {equation} 
      \vec p\cdot\vec q 
      = {1\over2} (\vec k + \vec k') ( \vec k - \vec k')               
      = {1\over2} ( \vec k ^{\,2} - \vec k^{\prime\,2}) = 0 
.\end {equation} 
The Jacobian of the transformation Eq.(\ref{eq:15}) 
is evaluated by means of the identities 
\begin {equation}
      \frac {\partial x} {\partial k _z}
      = { ( E _1 + k _z ) ( E _2 - k _z ) \over 
      E _1 E _2 \, ( E _1 + E _2 ) }
      \quad {\rm and } \quad 
      x(1-x)={(E_1+k_z)(E_2-k_z)\over (E_1+E _2)^2}
.\end {equation}
The auxiliary functions
\begin {equation}
      A ( \vec k , \vec k \,') 
      = \sqrt { {( E _ 1 + k _z ' )( E _ 2 - k _z ' )
      \over ( E _ 1 + k _z   )( E _ 2 - k _z   ) } }
      ,\qquad \quad {\rm and} \quad
      B ( \vec k , \vec k \,') 
      = \left( { m _r \over E _1 } + { m _r \over E _2 } \right)
\label{eq:24}\end {equation}
are useful for factorizing
\begin {equation}
      { m _r \ dx ^\prime \over \sqrt{x(1-x)\, x'(1-x')} } 
      = A ( \vec k , \vec k ' \,) 
      B ( \vec k , \vec k ' \,) \ dk _z
, \end {equation}
and to single out $A$ which is {\em not rotationally}
invariant. Both $A$ and $B$ are 
dimensionless and of order unity for sufficiently small 
momenta.
The cut-off function $ \Theta ' $, as introduced in
Eq.(\ref{eq:8}) to define a maximum transversal 
momentum restricts of course also three-momentum:
\begin {equation}
        \Theta (\vec k \,') :\qquad\qquad
        \vec k ^{\;2} \leq \left({\Lambda \over2}\right) ^2
        {\Lambda ^2 + 4 m _1 m _2 \over\Lambda ^2 + 
        ( m _1 + m _2) ^2}
\ . \end {equation}
The quark currents in Eqs.(\ref{eq:4}) or (\ref{eq:7}) can 
be evaluated with the Gordon decomposition \cite{BjD66}.
Since we work in the Lepage-Brodsky convention for the 
spinors \cite{BrP91} one has
\begin {eqnarray} 
      \bar u (k_1,\lambda_1)\,\gamma^0\,u(k'_1,\lambda'_1) 
     &=&  \left( E _1 + m _1  
     + { \vec p ^{\,2} - \vec q ^{\,2}/4\over( E _1+ m _1) } 
     + { \vec R \cdot\vec\sigma 
      \over(E_1+m_1) }\right)_{ \lambda_1,\lambda_1'}
, \\
      \bar u (k_1,\lambda_1)\,\vec\gamma\,u (k'_1,\lambda'_1) 
      &=& \left(2\vec p 
      - i\,\vec q\times\vec\sigma\right)_{\lambda_1,\lambda_1'}
, \\ {\rm with} \quad
      \vec R &=& i \vec q \times\vec p  
. \end {eqnarray} 
These expressions are simpler than usual, 
due to Eq.(\ref{eq:19}).
For the antiquark one must change the sign of both 
$\vec k$ and $\vec k'$, and replace the quark-spin matrix 
$\sigma$ by  $\tau$. 
The current term becomes then explicitly
\begin {eqnarray} 
      J ( \vec k , \vec k ' \,) 
      &=& {1\over 4 m_1 m_2} 
      \left[ \overline u (k_1,\lambda_1) 
      \,\gamma^\mu\,
      u(k_1^\prime,\lambda_1^\prime)\right] 
      \left[ \overline u (k_2,\lambda_2) 
      \,\gamma_\mu\,
      u(k_2^\prime,\lambda_2^\prime)\right] 
\\
      &=& C \left(  1 
      + {\vec p ^{\,2}-\vec q ^{\,2}/4\over( E _1+ m _1) ^2} 
      + {\vec R \cdot \vec \sigma \over ( E _2 + m _2) ^2 } 
        \right)  \left(  1 
      + {\vec p ^{\,2}-\vec q ^{\,2}/4\over( E _1+ m _1) ^2} 
      + {\vec R \cdot\vec\tau\over(E_2+m_2)^2}\right) 
\nonumber \\
       &+& C\left({2\vec p -i\,\vec q \times\vec\sigma
       \over E_1+m_1}\right) \left(
       {2\vec p -i\,\vec q\times\vec\tau\over E_2+m_2}\right)
, \label{eq:31}\\ 
       {\rm with} \quad   C 
       &=&{( E _1+ m _1)( E _2+ m _2)\over 4 m _1 m _2 } 
. \end {eqnarray} 
After $A$ and $B$ a third auxiliary 
function $C$ is introduced, which is also
dimensionless and of order unity.
As expected for a Lorentz scalar, $J$ is 
rotationally invariant.

Thus far, all quantities considered are of order unity
for sufficiently small momenta. The most important 
part of the interaction kernel 
\begin {equation} 
        \langle\vec k\vert U
        \vert \vec k^{\,\prime} \rangle =
        \Theta (\vec k ') 
        \, A (\vec k , \vec k ' )
        \, B (\vec k , \vec k ' )
        \, J (\vec k , \vec k ' )
        \ \widetilde V (\vec k , \vec k ' )
\label{eq:33}\end {equation}
is therefore the interaction proper
\begin {equation}
      \widetilde V ( \vec k , \vec k ' \,) 
      = -{1\over 2 \pi ^2} 
      {\alpha_s (Q^{2}) \over Q^{2} }
      = -{1\over 2 \pi ^2} 
      {\alpha_s (\vec q ^{\, 2}) \over \vec q ^{\, 2} }
. \end {equation}
It depends only on the three-momentum transfer. 

The front form is frame and boost invariant, as mentioned.
It is rotationally {\em co-} but not rotationally {\em in}variant,
particularly when the spatial rotations
are performed perpendicular to the $z$-axis. This aspect
is reflected in the appearance of the factor  $ A$
as defined in Eq.(\ref{eq:24}). The violation of rotational
invariance occurs however in such a form that it can be
absorbed into the wavefunction. If one inserts
\begin {equation}
        \phi(\vec k\,) = \langle\vec k\vert\psi\rangle\,
        \sqrt { {(E_1+k_z')(E_2-k_z')\over E_1E_2} }
\label{eq:35}\end {equation}
into Eq.(\ref{eq:16}), the factor $ A$ cancels 
in the new integral equation 
\begin {equation} 
        E\,\phi(\vec k) 
        = T(\vec k )\,\phi(\vec k)  
        +  \int \! d^3 \vec k ^\prime  
        \,\Theta (\vec k ') 
        \, B (\vec k , \vec k ^\prime)
        \, J  (\vec k , \vec k ^\prime)
        \,\widetilde V (\vec k , \vec k ^\prime)
        \ \phi(\vec k^\prime)  
\ .    \label{eq:36}\end {equation}
The kernel is now rotationally invariant.
Since no approximations have been made, the solutions 
of this equation, {\it mutatis mutandis}, are  identical with 
those obtained  from the original front-form integral 
equation, Eq.(\ref{eq:4}), but Eq.(\ref{eq:36}) is 
much easier to deal with.

\section{The retarded Schr\"odinger equation}\label{sec:4}

The front form of Hamiltonian dynamics \cite{Dir49} has
wonderful properties but it does not appeal strongly to our
intuition,  not even when it is transcribed  to instant form 
variables.  Thinking in terms of momentum-space  integral
equations is not always easy. The equations become more
transparent when Fourier transforming them to configuration
space and the corresponding Schr\"odinger form of quantum
mechanics.

We begin with rewriting  Eq.(\ref{eq:36}) conveniently as
\begin {equation} 
        E\,\phi(\vec k) = \int\!\! d^3 \vec k ^{\,\prime}\, 
        \widetilde H (\vec q , \vec p ) \ \phi ( \vec k \,')
\ .    \label{eq:37}\end {equation}
The kernel $\widetilde H$ is expressed in terms of the 
momentum transfer and its mean rather than by 
$ \vec k $ and $ \vec k '$. It is the Fourier transform of the
Schr\"odinger  Hamiltonian. To see that one multiplies the 
whole equation 
with  $ \exp ( i\vec k \cdot \vec x ) $ and integrates over 
$ d^3 \vec k $. Defining the Fourier transforms by
\begin {equation} 
        \psi( \vec x \, ) = \int\!\! d ^3 \vec k 
        \ {\rm e} ^{ i\vec k \cdot \vec x } \ \phi (\vec k \,' ) 
        \ ,\qquad {\rm and} \quad 
        H (\vec x , \vec p \, ) = \int\!\! d ^3 \vec q
        \ {\rm e} ^{ i\vec q \cdot \vec x } 
        \ \widetilde H ( \vec q , \vec p \,) 
\ ,    \end {equation}
one gets an eigenvalue equation of the Schr\"odinger 
type with a possibly non-local Hamiltonian 
\begin {equation}
        E \psi (\vec x ) = 
        H (\vec x ,\vec{\underline{p}}\,) \ \psi ( \vec x )
        \ , \qquad {\rm with} \quad 
        \vec{\underline{p}} \equiv -i \vec \nabla _x
\ .    \end {equation}
The momentum transfer $\vec q $ is Fourier conjugate to the
position $\vec x$ of the quark in the center-of-mass frame,
and  $\vec{\underline{p}}$ is the associated momentum
operator. This holds in general, but unfortunately one is 
unable to perform the Fourier transform explicitly with all
the square roots behind the energies $E_i$. 
The way out is, of course, to expand and to develop a 
systematic approximation scheme. We base it on the 
Lepage-Brodsky cut-off and choose $\Lambda$ such that 
\begin {equation}
        {\vec k ^{\;2} \over m _1 ^2 } \leq 1
\ ,     \end {equation}
for the lighter quark $m _1$. 
All square-roots are  expanded to first non-trivial order
\begin {equation}
        E _i \simeq 
        m _i + { \vec k ^{\, 2} \over 2 m _i } =
        m _i + { \vec p ^{\, 2} \over 2 m _i } 
                 + { \vec q ^{\, 2} \over 8 m _i } 
\ ,    \label {eq:41} \end {equation}
which is a semi-relativistic approximation. In the worst 
case it allows for relativistic velocities of the lighter particle 
up to $\vert\vec k \,\vert\sim m _1 $. 
The expansion of the various factors in the kernel of 
Eq.(\ref{eq:36}) yields  up to second order
\begin{eqnarray} 
        B 
        &=& 1 - {\vec p ^{\,2} \over 2 m _q ^2} -
        {\vec q ^{\,2} \over 8 m _q ^2 }
        \ ,\,\ \quad{\rm with}\quad 
        {1\over m _q ^2} = {1\over m _1+m _2}
        \left( {m _2\over m _1^2} + {m _1\over m _2^2} \right)
\ , \\ C 
        &=& 1 + {\vec p ^{\,2} \over  4 m _a ^2 } +
        {\vec q ^{\,2} \over 16 m _a ^2 }
        \ ,\quad{\rm with}\quad 
       {1\over m_a ^2} = {1\over m_1^2}+{1\over m_2^2} 
\ , \\ {\rm and}\ B J
        &=& 1 + {\vec p ^{\,2}\over 2 m _1 m _2}
        - {\vec q ^{\,2}\over 8 m _q ^2}
        - {(\vec\sigma \times \vec q\,) \cdot 
          (\vec\tau   \times \vec q\,) \over 4 m _1 m _2 }
       + { \vec\sigma \cdot \vec R \over 4 m _1 ^2}
       + { \vec\tau   \cdot \vec R \over 4 m _2 ^2}
       - { \vec S   \cdot \vec R \over   m _1 m _2}
\ ,    \end{eqnarray}
respectively. One should emphasize that the form of
$\widetilde V (\vec q)$ need not be known at this point,
since it does not depend on $\vec p$.
The total spin and  the kinetic energy,
\begin {equation}
       \vec S = {1\over2} (\vec\sigma + \vec\tau)
       \qquad{\rm and}\qquad 
       T =  {\vec k ^{\,2} \over 2 m _r}
\ ,\end {equation}
respectively, complete the definitions. Finally, one can 
conjecture that the wavefunction decays sufficiently fast, 
such that it acts itself like a cut-off. We therefore set 
$ \Theta (\vec k \,') = 1$.

The Hamiltonian operator in Schr\"odinger representation
becomes then straightforwardly
\begin {eqnarray} 
    H 
   &=&  { 1 \over 2 m _r } 
        \left (1 + { V (r)\over m _1 + m _2} \right )
        \vec{\underline{p}}^{\,2}  
     +  V (r) 
        + {\vec \nabla ^{2} V (r) \over 8 m _q ^2}
        +  {(\vec\sigma \times \vec \nabla) \cdot 
            (\vec\tau   \times \vec \nabla V ) 
             \over 4 m _1 m _2 }
\nonumber \\ 
       &+& {1\over r} {\partial V \over \partial r } \left (
           { \vec\sigma \cdot \vec L \over 4 m _1 ^2}
         + { \vec\tau   \cdot \vec L \over 4 m _2 ^2}
         - { \vec S   \cdot \vec L \over   m _1 m _2} \right )
\ , \end {eqnarray} 
with the usual angular momentum operator 
$ \vec L  =  \vec x \times \vec{\underline{p}}$.
Since the average potential is spherically symmetric, 
one uses
$ (\vec\sigma \times \vec \nabla) \cdot 
            (\vec\tau \times \vec \nabla V ) 
  = {2\over3} (\vec\sigma\cdot\vec\tau) \vec \nabla ^{2} V$ 
and $ \vec\sigma\cdot\vec\tau = 2 \vec S ^{\,2} - 3 $ 
to get 
\begin {eqnarray} 
        H  &=& { 1 \over 2 m _r } 
        \left (1 + { V (r)\over m _1 + m _2} \right )
        \vec{\underline{p}}^{\,2}  
    +  V (r) 
        -              {\vec\nabla ^{2} V \over 8 m _r ^2}
           \left( {3m _r\over m _1 + m _2}
           - {( m _1 - m _2)^2\over( m _1 + m _2)^2} \right)
\nonumber \\ 
       &+& {1\over r} {\partial V \over \partial r } \left (
           { \vec\sigma \cdot \vec L \over 4 m _1 ^2}
         + { \vec\tau   \cdot \vec L \over 4 m _2 ^2}
         - { \vec S   \cdot \vec L \over   m _1 m _2} \right )
        +  {\vec\nabla ^{2} V \over 3 m _1 m _2 } \vec S ^{2} 
\ . \label{eq:47}\end {eqnarray} 
Its structure is a direct consequence of gauge theory  
particularly QCD and holds for an arbitrary running coupling
constant $\alpha_s(Q^2)$. We emphasize particularly
that this structure was obtained from a fully covariant theory
\cite{Pau96}.
The statement could even  be stronger  without our inability 
to evaluate the Fourier transforms without expansions. 
Richardson's parametrization
of $\alpha_s(Q^2)$ yields the potential $V(\vec x)$
as given in Eq.(\ref{eq:11}), and thus 
\begin {equation}
     \vec \nabla ^{2} V (r) = \beta 
     \left ({2\kappa ^2 \over r } + 4\pi\delta(\vec x )\right )
\,   \qquad {\rm and} \quad
     {1\over r} {\partial V (r) \over \partial r } 
   = \beta \left ( {\kappa ^2 \over r} + {1 \over r ^3} \right )
\ ,   \end {equation} 
with $\beta = {8\pi\over 27} \simeq 0.93$.
If one works with QED, one sets $\kappa=0$ and chooses
the value $\beta\simeq 1/137$.

We now have reached our goal:
The retarded Schr\"odinger equation and its Hamiltonian 
have a wonderfully simple structure which can be interpreted
with ease. The average potential $V(\vec x)$ plays a different
role in  the different terms of the equation. In the first term
of Eq.(\ref{eq:47}), 
in the kinetic energy, it generates an effective mass
of the quark which depends on the relative position and 
which reflects the non-locality of the interaction.  
In the second term,  $V(\vec x)$ appears in its natural role 
as a potential energy. In the third term one observes 
$V(\vec x)$ as the analogue of the Darwin term.
In the remainder $V(\vec x)$ provides the coupling strength 
for the analogue of the fine and  hypefine interactions of 
atomic physics particularly the spin-orbit interaction. 
Contrary to common belief they exist not due to weak 
coupling but they appear also for strongly coupled QCD.

Finally, one must come back to the expansion scheme of
Eq.(\ref{eq:41}). Its validity cannot be jugded a priori, 
since the expansion is made under the integral. 
The omitted terms are of second order in $p^2/m^2$ 
for the lighter quark. Whether this is justified or not 
can be decided only a posteriori, by the expectation value
of the omitted next higher term
\begin {equation}
     \delta \equiv {1\over8} 
     \left ({\langle\vec p ^{\,2}\rangle\over m _i ^2}\right) ^2
\ .  \label {eq:49} \end {equation}
Only if $\delta$ is (very) small as compared to unity,
the expansion in Eq.(\ref{eq:41}) is justified. 
If it is comparable or larger than 1, the solution 
must be rejected,  and  another regime
of approximation must be found. Below, we shall meet
cases like that.
\section{Meson masses by parametric variation} 
\label{sec:5}

It will take some time and effort to work out all the many 
consequences of the integral equations,  Eq.(\ref{eq:4}) or
Eq.(\ref{eq:36}), or of the retarded Schr\"odinger equation
(\ref{eq:47}). In the sequel, we shall restrict ourselves 
to calculate only the ground-state masses of 
the pseudoscalar and  vector mesons. 
If one leaves aside the recently discovered top quark 
and restricts to 5 flavors ($u,d;s,c;b$), one has thus 30 
different physical mesons, since charge-conjugate hadrons 
have the same mass. 

One cannot calculate these masses, however, without 
knowing the quark mass parameters $ m _1$ and $ m _2$. 
These cannot be measured  in a model-independent way.
In the sequel, we shall adopt the point of view that they 
have to be determined consistently within each model, 
for the better or the worse. One has thus 5 mass parameters 
to account for 30 physical masses. Which ones should be 
selected to fit? There are 142~506 different possibilities 
to select 5 members from a set of 30, and we have to make 
a choice: We choose the five pure $q\bar q$-pairs. 
Even that is not unique: Shall one take the pseudoscalar 
or the vector mesons? We shall do both. 

Of course, one runs into the problem of  the chiral 
composition of the physical hadrons. In order to avoid that
in the very crude estimate below we shall substitute all
physical mesons by pure $q\bar q$-pairs~-- {\it by fiat}. 
We shall thus replace the `pions', for example, by 
`quasi-pions' with the same physical mass. 
The  $u\bar u$-, $u\bar d$-, $d\bar u$-, or 
$d\bar d$\--eigenstates shall be identified with the
quasi-$\pi^0$,  quasi-$\pi^+$, quasi-$\pi^-$, 
or the quasi-$\eta$, and so on. 
This simplification will be revoked in future work and is, 
by no means, a compelling part of the model.

Our problems lie in another ball park. One should not deal
head-on with the full complexity of the integral equations, or
of the retarded Schr\"odinger equation.  Which part of the
Hamiltonian should one select in a first  assault? Some 
help is gained by the rather unique property of the light-cone
Hamiltonian: kinetic and interaction energy are additive. 
One can select always an `interesting part' $H_0$, and 
check {\it e posteriori}, by calculating the expectation 
value of $\Delta H = H - H _0$,  
whether the selection makes sense, or not. 
Since the scalar and pseudoscalar mesons have only 
little orbital excitations, {\it i.e.} are primarily $s$-waves, 
one can disregard the spin-orbit part and choose first
\begin{equation} 
      H _0  \sim 
      \left( 1+{V(r)\over m_1+m_2} \right)
      {\vec p ^{\,2} \over 2 m_r}  +  V (r) 
     + {\vec\nabla^{2} V\over 3 m_1 m_2 }\vec S^2 
      - {\vec\nabla^{2} V\over 8 m_r ^2}
     \left( {3m_r\over m_1 + m_2}
     - {( m_1 - m_2)^2\over( m_1 + m_2)^2} \right)
\ . \end {equation} 
Even that looks to complicated for a start-up. 
We therefore select those terms which have turned out 
to be important in the past, namely the central potential
and the triplet-interaction mediated by the total spin. 
Omitting the effective mass and the Darwin term, our 
choice is therefore \begin {equation} 
  H _0  = { \vec p ^{\,2} \over 2 m _r } + V (r) 
     + {2 \kappa \over 3 m _1 m _2 } {\vec S ^{2} \over r} 
     = { \vec p ^{\,2} \over 2 m _r } 
     - {\beta\over r }
     + {2 \kappa ^2\over 3 m _1 m _2 } {\vec S ^{2} \over r} 
     +  \beta \kappa ^2 r 
\ . \label{eq:52}\end {equation} 
Working in a spinorial representation which diagonalizes 
$ S_z$  and $ \vec S ^2$, we replace the latter  by the
eigenvalue $ S _e = S( S + 1)$ and take  0 or 2 for the 
singlet or the triplet, respectively. 

How does the wavefunction for the lowest state look like? 
For a pure Coulomb potential the solution has the form
\begin {equation}
  \psi (\vec x ) 
    = {1\over \sqrt{\pi} } \lambda ^{{3\over2}} 
      \, {\rm e} ^{-\lambda\, r} 
\ .  \end {equation}  
Omitting the Coulomb part, a linear potential can be solved 
in terms of Airy functions and its integral transforms
\cite{PaB79}. If one has both, one will have some mixture 
of the two. But for the present start-up check even that 
requires too much effort. 

\begin {table}[t]
\begin {center}
\caption{\label{tab:1} \sl
   The flavor quark masses in MeV, 
   as obtained from a fit to Eq.(\protect\ref{eq:57}). 
   The first row refers to a fit for the singlets, 
   the second to the one for the triplet.
}\vspace{0.5em}
\begin {tabular}  {|c||c|c|c|c|c|}
\hline 
                  {\rule[-3ex]{0ex}{7ex}}
                  {\rm Flavor\ mass} 
                  & u & d & s & c & b 
\\ \hline \hline   
                  {\rule[-2ex]{0ex}{4ex}}
                  {\rm From\ fit\ to\ $0^-$} 
               \  & 2.3   & 155.6 & 430.6 & 1642.3 & 5330.8 
 \\ 
                  {\rule[-2ex]{0ex}{5ex}}
                  {\rm From\ fit\ to\ $1^-$} 
               \  & 222.8 & 236.2 & 427.2 & 1701.3 & 5328.2 
 \\ \hline 
\end {tabular}
\end {center}
\end {table}
Rather shall we pursue a variational approach and choose 
Eq.(\ref{eq:52}) as a one-parameter family ($\lambda$).
One could take also  harmonic oscillator states
\cite{BrS94,Sch94}, but with Eq.(\ref{eq:52})
the expectation values are particularly simple:
\begin {equation}
     \langle\psi\vert\, \vec p ^{\,2} \,\vert\psi\rangle 
    = \lambda ^2 
\ , \qquad 
     \langle\psi\vert\, {1\over r} \,\vert\psi\rangle 
    = \lambda 
\ , \qquad {\rm and} \quad
     \langle\psi\vert\, r \,\vert\psi\rangle 
    = {3\over 2 \lambda }
\ . \end {equation}  
These are all one needs for calculating the expectation
value of the energy 
\begin {equation} 
   \overline E = \langle\psi\vert H _0 \vert\psi\rangle 
     = { \lambda ^2 \over 2 m _r } 
     - \beta\lambda 
     + {2 S _e \kappa ^2 \over 3 m _1 m _2 } \,\lambda 
     +  {3\beta \kappa ^2 \over 2 } {1\over\lambda} 
\ .  \label{eq:54}\end {equation} 
Since we deal only with ground states, we are not in 
conflict with the statement that the wavefunction cannot 
be purely Coulombic.
For the pure Coulomb case, the 2$S$- and 1$P$-states 
would be degenerate and the respective ratio 
$|\psi _{2S}(0)| ^2/|\psi _{1S}(0)| ^2 = 0.125$ would 
disagree with the experimental values $\simeq 0.63$ and 
$\simeq 0.50$ for charmonium and bottomium \cite{QuR79}.

We aim at calculating the total invariant mass 
of the hadrons and  return to the light-cone Hamiltonian 
$ H_{LC} = M^2$, {\it i.e.} to  
$ M ^2 = ( m_1+m_2)^2+2(m_1+m_2)\overline E$.
For equal masses $m_1=m_2= m$ one preferably 
expresses the variational equation (\ref{eq:54}) 
in units of the fixed QCD scale $\kappa$, introducing 
the dimensionless variables
\begin {equation} 
     s = \frac{\lambda}{\kappa} 
\ , \qquad
   \xi = \frac{m}{\kappa} 
\ , \qquad {\rm and} \quad
     W = \left( \frac { M } {2\kappa} \right) ^2 
\ . \end {equation}   
The variational  equation (\ref{eq:54}) reduces then simply to 
\begin {equation}
  W (s;\xi) = s ^2 
        - \left (\beta\xi - {2 S _e \beta\over3\xi} \right ) s
        + \xi ^2 
        + {3\beta\xi\over 2} \,{1\over s} 
\ .  \end {equation}
We must vary $\lambda$, thus $s$, such that the energy 
is stationary, 
\begin {equation} 
      {\partial W (s;\xi) \over \partial s } 
       \bigg\vert _{ s = s ^\star (\xi)} = 0
      \ ,\qquad {\rm thus} \quad 
      \left( { M \over 2\kappa} \right) ^2 
  =   W \left ( s ^\star (\xi) \right) = W ^\star (\xi) 
\ ,  \label{eq:57}\end {equation}
at fixed values of the parameters ($\xi,\beta, S _e$). 
This leads to a cubic equation in $ s $ which can be 
solved analytically in terms of Cardano's formula. 
In special cases they can well be approximated by a 
quadratic equation, namely when 
$s^\star \gg 1$ or when $ s ^\star \ll 1$.
We got accustomed to refer to these two regimes 
as the Bohr  and the string regime, respectively. 
In the Bohr regime the Coulomb potential dominates the 
solution and the linear string potential provides a correction.
In the string regime the linear string potential dominates, 
with the Coulomb potential giving a correction.
Solutions in the string regime, however, imply that the 
ratio $ \langle\vec p^{\,2}\rangle/m^2 = \lambda^2/m^2$
becomes so large that one is in conflict with the 
validity condition Eq.(\ref{eq:49}).

\begin {table} [t]
\begin {center}
\caption{\label{tab:2} \sl 
    The validity check.
    The first row displays the values of $(\lambda/m)^2 $ 
    as obtained from the mass fit to the singlets, 
    the second row those from the mass fit to the triplets.
    If the mean momentum is comparable to the mass, or
    larger, the solution has to be rejected. 
    The extremely large value for the $u$-quark in 
    pseudoscalar fit gives a good example for such a case. 
}\vspace {0.5em}
\begin {tabular}   {|c||c|c|c|c|c|}
\hline 
                  {\rule[-2ex]{0ex}{6ex}}
                  $\displaystyle {\lambda^2\over m^2}$ 
                  & u & d & s & c & b 
\\ \hline \hline   
                  {\rule[-2ex]{0ex}{5ex}}
                  {\rm For\ fit\ to\ $0^-$} 
                \ & 293  & 1.45 & 0.53 & 0.25 & 0.22 
 \\ 
                  {\rule[-3ex]{0ex}{6ex}}
                  {\rm For\ fit\ to\ $1^-$} 
                \ & 0.79 & 0.76 & 0.48 & 0.25 & 0.22 
 \\ \hline 
\end {tabular}
\end {center}
\end {table}
Rather than to display explicitly the straightforward but 
cumbersome formalism, we present the analytical results 
in the graphical form of Figure~\ref{fig:3}.
The total mass $ M = 2 \kappa \sqrt { W ^\star} $ is almost 
linear in the quark mass, with small but significant 
deviations. In line with expectation, the hyperfine splitting 
decreases with increasing quark mass.
Less expected was that the splitting increases so strongly 
with decreasing quark mass.
For very small quark masses, the triplet mass starts off at a
finite and almost constant value. The singlet mass takes off
from zero like a square root but unfortunately not linearly as
required by the soft pion theorems.  Determining  the 
$u$-mass by fitting to the quasi-pion gives a value close
to the `current mass', see Table~\ref{tab:1}. The resulting 
$s\ll1$, see Tables~\ref{tab:1} and \ref{tab:2}, implies 
the ultra-relativistic string regime and that the  validity 
condition is badly violated. 
The retarded Schr\"odinger equation with its 
semi-relativistic approximation scheme is thus not 
appropriate for describing quasi-pions.
For the $\eta$ and the $\eta'$, the scaling variable $s$ 
is of order unity, while for the quasi-$\eta _c$ one 
definitely is in the Bohr regime.
Here the masses are similar or close to what is 
refered to as the `constituent-quark' mass. 

Since singlet and triplet are so close for $s\gg 1$, one fits
the quark masses preferentially with the vector mesons.
In the lack of empirical data we have set 
$ M _{\eta _b} = M _{\Upsilon} $, which should be of minor
importance in the present model, see Figure~\ref{fig:3}. 
The flavor masses are now in close agreement with the 
constituent quark masses, see Table~\ref{tab:1}, and
the smallness condition is satisfied better.
see Table~\ref{tab:2}. 

\begin {table} 
\begin {center}
\caption [BigTableII] {\label{tab:4} \sl 
    The masses of $q\bar q$--hadrons are 
    compared with experimental values.
    The flavor quark masses used are inserted 
    in column 2
    and come from a fit to the vector mesons.
    The first line within each box refers to the 
    hadronic symbol of the meson; 
    the second line gives the calculated (measured) 
    vector mass in MeV;
    the third line accounts for the calculated (measured) 
    pseudoscalar mass in MeV, 
    and finally the fourth line specifies 
    the hadronic symbol of the pseudoscalar meson.
} \vspace{1mm}
\begin {tabular}  {|c|c||c|c|c|c|c|}
\hline
   {\rule[-3ex]{0ex}{7ex}}                   
   \qquad { }\qquad     & $\bf m_q$          & $\overline{\bf u}$ 
   & $\overline{\bf d}$ & $\overline{\bf s}$ & $\overline{\bf c}$ 
   & $\overline{\bf b}$ 
\\ \hline \hline
   & & $\rho^{0}$ & $\rho^{+}$ & $K^{*+}$ 
     & $\overline{D}^{*0}$ & $B^{*+}$ \\
   & & 768(768) & 773(768) & 910(892) & 2110(2007) & 5712(5325) \\    
   {\rule[-0.2ex]{0ex}{0.8ex}} 
   \bf u & 222.8 & & & & & \\ 
   & & 714(135) & & & & \\
   & & $\pi^{0}$ & & & & 
\\ \hline
   & & & $\omega$ & $K^{*0}$ & $D^{*-}$ & $B^{*0}$ \\
   & & & 782(782) & 914(896) & 2109(2010) & 5709(5325) \\
   {\rule[-0.2ex]{0ex}{0.8ex}} 
   \bf d & 236.2 & & & & & \\
   & & 658(140) & 668(549) & & & \\
   & & $\pi^{-}$ & $\eta$   & & & 
\\ \hline
   & & & & $\phi$ & $D_s^{*-}$ & $B^{*0}_s$ \\
   & & & & 1019(1019) & 2156(2110) & 5735(~---~) \\ 
   {\rule[-0.2ex]{0ex}{0.8ex}} 
   \bf s & 427.2 & & & & & \\ 
   & & 825(494) & 831(498) & 953(958) & & \\
   & & $K^{-}$ & $\overline{K}^{0}$ & $\eta^{'}$ & & 
\\ \hline
   & & & & & $J/\psi$ & $B^{*+}_c$ \\ 
   & & & & & 3097(3097) & 6502(~---~) \\
   {\rule[-0.2ex]{0ex}{0.8ex}} 
   \bf c & 1701.3 & & & & & \\
   & & 2079(1865) & 2078(1869) & 2131(1969) & 3082(2980) & \\
   & & $D^{0}$ & $D^{+}$ & $D_s^{+}$ & $\eta_c$ & 
\\ \hline
   & & & & & & $\Upsilon$ \\
   & & & & & & 9460(9460) \\
   {\rule[-0.2ex]{0ex}{0.8ex}} 
   \bf b & 5328.2 & & & & & \\
   & & 5701(5278) & 5698(5279) & 5726(5375) 
                  & 6495(~---~) & 9455(~---~) \\
   & & $B^{-}$ & $\overline{B}^{0}$ & $\overline{B}^0_s$ 
     & $B^-_c$ & $\eta_b$ 
\\ \hline   
\end {tabular}
\end {center}
\end {table}
Having determined the quark masses one has exhausted 
all freedom in the model. We now ask:
How well agree the remaining 25 meson masses with
experiment?~-- The formal procedure runs quite
analogously, except that it is  now easier. With the masses 
fixed, one varies $\lambda$ separately for each flavor 
composition subject to  Eq.(\ref{eq:54}).
The results are compiled in Table~\ref{tab:4} and 
compared with the experimental values to the extent 
the latter are known. 
The present model predicts for example
\begin {equation}
    M (B _c ^{\pm})  = 6495 \,{\rm MeV}
    \ , \qquad 
    M (B _c ^{*\pm}) = 6502 \,{\rm MeV} 
\ . \end {equation}  
By and large the agreement is remarkably good.
The heavy meson masses are reproduced quite well,  but
the agreement is not quantitative everywhere, particularly
not for those hadrons with one light quark ($u$ or $d$). 
In judging this agreement one should keep in mind, (1) that
such a table, in which all hadrons have been calculated
from one and the same model, has hitherto not been 
prepared; and (2) that the light mesons like the quasi-pions 
should not be calculated by a crude potential model 
like the retarded Schr\"odinger equation. The smallness
condition actually tells us that these hadrons probably are 
systems in which the constituents move highly relativistically. 
Thus far, we have at hand no simple paradigms for such
a kinematic situation. Solving directly the momentum-space 
integral equations might therefore be the only way.

All in all, with all due respect to the work with potential models
 and with lattice gauge theory, the agreement between 
the empirical data and the present first attempt to relate them
on trial and error  to an effective,  QCD-inspired Hamiltonian is
in fact not so much worse; particularly in view of the absence
of any  free parameters and the simplicity  of the approach. 
No doubt, the various simplifications can  be improved  
in the future.

\section {Summary and Conclusions}
\label{sec:6}

The full many-body front-form Hamiltonian,
evaluated for QCD in the light-cone gauge $A^+ = 0$, 
had been reduced  in preceeding work \cite{Pau96} 
to a manifestly gauge invariant  effective Hamiltonian 
which acts only in the space of one quark and one antiquark.
Particularly, no Tamm-Dancoff Fock-space truncations
had to be made, nor was it necessary to rely on perturbation
theory by assuming a small coupling constant.
The present work is motivated by the question why and how
such a simple structure like the resulting integral equation
in the transversal momenta and the longitudinal momentum
fraction can account for the complexities of
hadronic phenomenology. In particular we have wondered
to which extent one can understand the masses of the 
pseudoscalar and vector  mesons with no other input than the
flavor quark masses of the constituent quarks.

In this first study of such a structure, which actually was
preceeding \cite{Mer94,Pau93} the more rigorous derivation
\cite{Pau96}, we replace the running coupling constant, 
which absorbes the many-body amplitudes of the full theory 
in a well  defined way, by the suitably adjusted 
phenomenological version of Richardson \cite{Ric79}. 
At the least, the latter interpolates smoothly between 
asymptotic freedom and infrared slavery.
Its only free parameter is fixed by a fit to the strong coupling
constant at the Z mass.

For the future, we have in mind mainly two
improvements: (1) the explicit  calculation of the
running coupling constant by a renormalization group
analysis, and (2) an explicit solution of the integral equation 
in light-cone variables Eq.(\ref{eq:4}).
It should be applied to mesons whose constituent quarks 
havevery different masses, 
such that the structure functions including the contributions
from higher Fock states can be calculated from a 
covariant theory. This could be done in such a way that
the relation to the existing phenomenological models
can be seen explicitly.

With these future applications in mind, we have
not hesitated to perform in Section~3 a number of basically 
trivial and straightforward calculation and to 
transcribe the front-form integral equation into the intuitively
easier accessible form of usual momenta. The major impact 
of very different quark masses is then absorbed into the
familiar reduced mass, and all integration variables have the
same domain of validity. This is not unimportant for the
practitioner who actually wants to get out numbers from his
theory. This virtue does not seem to be common place
anymore, unfortunately. As a wonderful and not intended side
effect, it turns out that the rotationally only
covariant equation on the light-cone can be transformed 
into a rotationally invariant  integral equation
Eq.(\ref{eq:36}) in usual momentum space. All factors
which seem to violate strict rotational invariance can be
absorbed into the wavefunction, Eq.(\ref{eq:35}). 
One looks forward to see numerical solutions to these
equations. 

But in our aim to relate the basically exact formalism 
with its connection to Lagrangian QCD to the usual 
configuration space where our intuition is at home, 
we went a step further and tried to Fourier transform 
the integral equations. We have been unable to do this,
by formal mathematical reasons. Rather we had to discourse
to an approximation to which we refer to as semi-relativistic.
The resulting retarded Schr\"odinger equation 
Eq.(\ref{eq:47}) has the amazing property of looking like 
a conventional Schr\"odinger equation with 
velocity-dependent interactions and still being
a fully covariant equation. It should be obvious that the
transition from the front form to the usual instant form 
momenta  and the subsequent Fourier transform to
configuration space does not change the basic feature of the
light-cone Hamiltonian to be manifestly frame independent.
Would one be able to perform the necessary Fourier
transforms in closed form, this statement could be phrased
even more rigorously.

One should emphasize that the retarded Schr\"odinger 
equation has no free parameter, since coupling constant 
and quark masses have to be determined from the experiment.
Fitting the 5 quark flavor masses to the 5 $q\bar q$-vector 
mesons exhausts all  freedom. 
The rest is structure: The 25 remaining pseudoscalar and
vector masses are then predicted and presented in 
Table~\ref{tab:4}. In comparison with the experimental data,
they are not much worse than those from conventional 
phenomenological models \cite {QuR79,GoI85}, or from 
heavy quark symmetry \cite {Neu93}, or even from lattice 
gauge calculations \cite {Mac93,BCS94,Wei94}, in particular
when keeping in mind the very rough and simple
approximations applied. The pions are reproduced more than
poorly and remain mysterious particles like in every other 
model not specially designed for them. The mesons with
one light quark do not yet meet the tough standard of the 
phenomenological models. The latter two aspects are 
possibly related to each other.

Conclusion: If such a poor model can do so well
one must be on the right track. It seems that the front form
Hamiltonian approach applied to quantum chromodynamics 
has made a big step forward.  
Intensified efforts are justified.

\section{Acknowledgement}
HCP thanks Stanley~J.~Brodsky for the many discussions and 
exchange of ideas over all those ten years particularly 
for his patience in listening to the ideas still vague 
at the time of the Kyffh\"auser meeting \cite{Pau93}.
``Of course'', he said, ``that's what Richardson did.''~-- In
the final phase of writing-up the content of the master 
thesis \cite{Mer94} we got to knowledge on similar ideas 
by Zhang \cite{Zha95}.
The authors appreciate the expertise of Gernot Vogt in 
preparing Figure~1.

\end{document}